\documentclass[12pt, a4paper]{article}

\usepackage[centertags]{amsmath}
\usepackage{amsfonts}
\usepackage{amssymb}
\usepackage{amsthm}
\usepackage{graphicx}

\begin{document}

\title{Models of {DNA} denaturation dynamics: universal properties}
\author{M. Baiesi$^\dag$, E. Carlon$^\ddag$,\\
$^\dag$Dipartimento di Fisica e Astronomia, Universit\`a di Padova, and\\
Sezione INFN di Padova,\\ Via Marzolo 8, I-35131, Padova, Italy\\
$^\ddag$Institute for Theoretical Physics, KULeuven, \\
Celestijnenlaan 200D, B-3001 Leuven, Belgium}

\maketitle

\begin{abstract}
We briefly review some of the models used to describe DNA denaturation
dynamics, focusing on the value of the dynamical exponent $z$, which
governs the scaling of the characteristic time $\tau\sim L^z$ as a
function of the sequence length $L$.  The models contain different
degrees of simplifications, in particular sometimes they do not include
a description for helical entanglement: we discuss how this aspect
influences the value of $z$, which ranges from $z=0$ to $z \approx 3.3$.
Connections with experiments are also mentioned.\footnote{An 
edited version of this manuscript was published in:\\ Markov Processes Relat. Fields 19, 569-576 (2013)}
\end{abstract}

\section{Introduction}

Helical structures, also in a double helical form, are ubiquitous
in nature.  The most famous example is of course DNA. The
main reason for which nature has selected a double helical form
for the molecule which stores all genetic information is probably
its stability~\cite{Alberts02:_book}.  In fact, the twisting of the
strands around each other adds cohesion to the whole molecule because
the geometrical entanglement helps the base pairing interactions to
prevent thermal fluctuations from opening the DNA at random.

The entanglement of the two strands around each other is expected to have
a strong influence on the denaturation dynamics, and in the characteristic
times needed to separate the strands from one another. However, several models
of DNA denaturation do not explicitly take helical degrees of freedom
into account.  In this paper we review some models of DNA denaturation
dynamics, discussing their advantages and shortcomings.  Our focus
is the asymptotic scaling of the characteristic time of denaturation
$\tau$ as the function of the sequence length, $L$, which is expected to
behave asymptotically as \[ \tau\sim L^z \] for lengths $L\gg \ell_p$,
where $\ell_p$ is the persistence length.  In the previous equation $z$
defines the dynamical exponent. As we will see, values found for $z$
in the literature are spread from $z=0$ up to $z=3.3$, depending mostly
on whether the helical degrees of freedom are taken into account.
It is natural to expect that 
the time necessary to disentangle
a double helix scales with its length. 
In living cells a class of enzymes called topoisomerases~\cite{Dean85:_topois}
cut and rejoin portions of DNA to remove unwanted knots, linking,
or excessive twist. In an in vitro system where topoisomerases
are absent the two strands have to rotate around each other to loose
their twist and separate from each other.

Here we consider a laboratory situation where double-stranded DNA
is immersed in an environment that facilitates its denaturation,
such as high temperature aqueous solution and/or suitable  ionic
conditions.  Experiments showed that the fraction of molten DNA increases
for increasing temperature~\cite{Wartell85}.  Denaturation in this case
is an entropy-driven process, as the energy that was crucial to bind
the two DNA strands is overwhelmed by the entropic gain that the two
strands achieve by moving away from each other. The passage from the
initial ordered state to the final coil state goes through a sequence
of intermediate states that should be understood.

To date there are several simplified models that try to capture the
salient features of the DNA melting dynamics.  The simplifications, in
some cases very strong, are necessary to reduce the huge number of degrees
of freedom of a long DNA duplex to a manageable one in simulations, or to
have models that are simple enough to be treated analytically. However,
sometimes the simplifications could be meaningful in a restricted context,
and might be excessive for the problem considered.
So far, experiments of denaturation dynamics~\cite{Altan03} were
restricted to short sequences (e.g. $\approx 20$ base pairs). Thus the 
determination of the dynamical exponent $z$ from experiments remains
an open issue.
In any case, in the following
we present some arguments about the merits and shortcomings of some of
the models proposed in the literature for describing the dynamics of DNA
denaturation. The two classes of models we mainly consider are the Poland-Scheraga
(PS) model~\cite{PS66,Kunz07,Baiesi09} (or other simple directed models
of polymers~\cite{Maren02}) and three-dimensional
simplified models of polymers~\cite{Baumg86,Baiesi10}.
We also briefly discuss results obtained with the 
Peyrard-Bishop model~\cite{PB89,DPB93,vanErp12}.

\section{Poland-Scheraga and related models}

\begin{figure}[t!]
\begin{center}
\includegraphics[width=7.2cm]{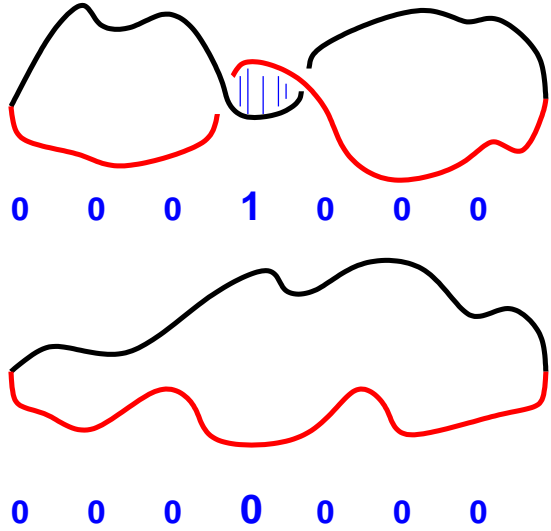}
\end{center}
\caption{Sketch of PS configurations with $L=7$ sites: (top) ``0001000''
represents two bubbles separated by a helix; (bottom) a single bubble
``0000000''. A ``Glauber'' move applied to the ``1'' of the first
configuration would yield the second one, but it would imply a nonlocal
rotation of either the left or the right bubble.}
\label{fig:PS}
\end{figure}

A state in the PS model~\cite{PS66} consists in a string of $L$ bits, either $0$ or
$1$, and hence it is evidently a strong reduction of the DNA degrees of
freedom, see Fig.~\ref{fig:PS}.  Each ``$1$" represents a bound portion,
which for us is conveniently identified with a full helical turn of
$\approx 10$ base pairs, with an associated a Boltzmann
weight $q = \exp[-\epsilon/k_B T]$.  
A purely entropic weight is associated to open bases, represented by $0$'s.
Each string of $\ell$ consecutive $0$'s represents a DNA bubble of
length $2\ell$ (each strand contributes with $\ell$ steps to one half
of the bubble).  Say that a configuration contains $m$ of such bubbles:
each bubble $i\in [1,m]$ of length $2l_i$ carries an entropic weight
\[
w_i\simeq A s^{2 \ell_i} \ell_i^{-c}
\]
as derived from the equilibrium statistical mechanics of
polymers~\cite{Kafri00}.
Here $s$ is the ``fugacity" per unit of length, while the factor
$\ell_i^{-c}$ describes a universal power-law correction with
$c \approx 2.14$
if self-avoidance between all polymer
segments is taken into account~\cite{Kafri00,Baiesi03}.
The constant $A$ in this
case includes the cost of starting a bubble in the DNA.  Thus the full
weight of a configuration with $b$ bound pairs and $m$ bubbles is
\[
W = q^{b}\times w_1 \times w_2 \times \cdots \times w_m  
\]
with the constraint $L=b+\ell_1+\ell_2+\ldots+\ell_m$.

Originally the model was developed
for open linear DNA, without constraints on the number and dimension
of bubbles. The picture is different if the twist between the two
strands is conserved, as in a PS model of DNA loops~\cite{Rudnick02},
or if plectonemic structures storing helicity are included in the
model~\cite{Alkan12}. These works show that the effect of helical
constraints is relevant for thermodynamical properties. Let us now see
what are the differences between dynamical properties of models with or
without helices.

Consider two configurations $x$ and $y$ with equilibrium weights 
$W_x$ and $W_y$. In simulations,
detailed balance in the dynamics is preserved if the rate $k(x\to y)$
for going from $x$ to $y$ and its
reverse $k(y\to x)$ satisfy $W_x\, k(x\to y) = W_y\, k(y\to x)$.
This can be achieved for example with a Metropolis acceptance rule.
There are several choices of dynamics, depending on the
updating rules for a given configuration. If one applies a Glauber-like
scheme with local moves changing a $0$ into a $1$ or viceversa (an
example is in Fig.~\ref{fig:PS}), the denaturation time $\tau\sim 1$,
i.e. $z=0$~\cite{Kunz07}. But this dynamics does not take the entanglement
of the two strands into account: after the  breaking the bonds the two
two strands still need to disentangle.
Another possibility for the local updates is a Kawasaki-like strategy
$01 \leftrightarrow 10$ preserving the amount of entanglement (number
of $1$'s) in the bulk~\cite{Baiesi09}.  Such helical entanglement is
dissipated only at the boundaries (site $1$ and site $L$), which is the
typical transition when the setup is such that the weights $q < s^2$.
If this scheme is applied, starting with bound configurations from
a low temperature regime, the denaturation time $\tau\sim L^z$ with
$z\simeq 3$. Moreover, the scaling of the number of bubbles reveals
a maximum for another intermediate timescale $\tau_1\sim L^{2.15}$.
A possible explanation for the large $z\simeq 3$ value was also
provided in~\cite{Baiesi09}.  The latter approach, trying to include
entanglement effects due to the double-helical initial constraint,
yields typical timescales that resemble those of another polymeric
dynamics where entanglement is relevant, namely reptation of polymers
in dense melts~\cite{Carlon01:_RDmodel,Paul91:_rept_simul}.

\begin{figure}
\begin{center}
\includegraphics[width=8cm]{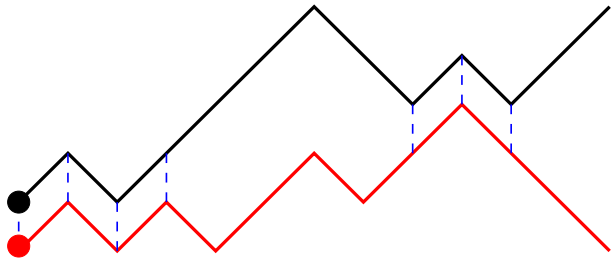}
\end{center}
\caption{Sketch of the model proposed in~\cite{Maren02}:  two 
interacting directed walks on a tilted square lattice.}
\label{fig:Maren}
\end{figure}

The shortcomings of the PS approach are its strong dependence on the choice
of dynamical rules, the fact that by definition only configurations
in the form paired-unpaired with respect to the DNA sequence are allowed,
which is a rough coarse-grained representation of the actual configuration
of the molecule during denaturation.
%

Besides the PS model, also models of directed polymers were
proposed. They are simple enough that an analytical treatment is
possible. An example is the model sketched in~\ref{fig:Maren}, proposed
by Marenduzzo et al~\cite{Maren02}. A pair of directed polymers starting
from neighboring sites on a tilted square lattice are mutually avoiding
and gain a unit of energy each time they are close to each other (dashed
lines in the figure).  One can also apply a force pulling away the two
right extremities, but the case without force is what we are interested in
here. Compared to the PS model, one can see that the bound segments have
some entropy, since there are two directions for every step. Moreover,
the entropy of the bubbles is not assumed. However, this model still
misses the key ingredient of the helical degrees of freedom, and hence
again its dynamics resembles more a polymer desorption than an unwinding.
The exponent found for the case without forces (and with bubbles)
was $z=4/3$.

\section{Simulations of lattice and off-lattice polymers}

The previous classes of models, although quite efficient to simulate
and study, still contain some approximations. For instance, 
the three-dimensional entanglement of the two chains is neglected.
Computer simulations of interacting self-avoiding polymers allow to
study the dynamics of denaturations of chains of about $L \approx 10^3$
monomers without resorting to uncontrolled approximations.  In these
simulations, hydrodynamics effects are usually neglected and
the polymer configurations are updated following a sequential local
update using detailed balance dynamics. Particularly interesting are
lattice polymers in which the updates consist in corner and end- flips,
which can be very efficiently  implemented and
correspond to Rouse dynamics~\cite{doi89}.  These types of simulations
were recently employed to study several aspects of the denaturation
dynamics~\cite{ferr11}, in the case of absence of bubbles and without
winding of the two strands (the two polymer strands are paired as in a
``zipper").  In this system the denaturation times scale with a dynamical
exponent $z=1$, whereas renaturation dynamics is ``anomalous" with $z
\approx 1.4$, an exponent also found in simulations of translocation
dynamics~\cite{vock08,luo09}.

The first study of disentanglement of double helical polymers dates
back to Baumg\"artner and Muthukumar~\cite{Baumg86} about thirty years
ago. Despite generating the initial double helix with monomers on a cubic
lattice, they then applied off-lattice corner flips. The steric constraint
between monomer was preserved thus with repulsive potentials between
them that was strong enough to prevent strand passages.  The results
they found included a large value $z\simeq 3.3$.  These simulations were
limited to relatively short chains, due to the limited computational
power available at that time.

\begin{figure}[t!]
\begin{center}
\includegraphics[width=11cm]{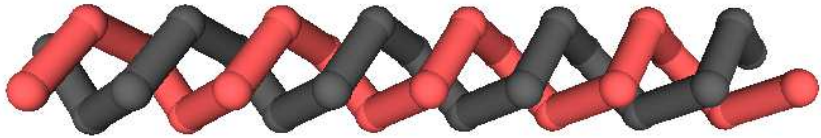}
\end{center}
\caption{Double helical SAWs on the fcc lattice, with $N=16$ steps per
strand, and $4$ sites per helical turn.}
\label{fig:fcc}
\end{figure}

Recently a simulation of a self-avoiding walk (SAW) variant on
the face-centered cubic (fcc) lattice yielded new results for long
walks~\cite{Baiesi10}. Again, starting from a double helical shape
(see Fig~\ref{fig:fcc}) and letting the two walks to move via local
rearrangements, it was monitored the minimal mutual distance $d$
between each of the monomers of one SAW from each of the monomers of
the other SAW.  Its square stays equal to the fcc lattice unit $d^2=2$
for a long time, until it starts to fluctuate to higher and higher
values. The time when  $d^2>10$ or $d^2>20$ for the first time, denoting
the disentanglement of the chains, was used to define the denaturation
time. Its average in both cases scales as $\tau\sim L^{2.57(3)}$,
thus there is yet another candidate exponent $z=2.57$ for describing
the disentangling process of two polymers prepared in a double-helical
conformation.

The discrepancy between the more recent $z\simeq 2.57$ and previous
$z\simeq 3.3$ is probably due to the difference in asymptoticity of
the respective simulations. Since numerical results can be plagued by
strong finite size effects, it is of course possible that $z\simeq 2.57$
is also not an asymptotic value yet.  The equilibrium properties of
winding angles in a system composed by a polymer attached to an end to an
infinite straight rod are characterized by ratios of angles to logarithms
of the chain length~\cite{Walter11}.  We are introducing these systems
because a polymer wrapped as a helix around a bar can be considered a
good surrogate of a double helical structure.  The presence of $\log L$
in the statistics of this systems warns us that the simple power-law
assumption $\tau \sim L^z$ might need some logarithmic corrections.
This problem is currently under investigation~\cite{Walter12}.

\section{Peyrard-Bishop model}

Another classical model for describing DNA denaturation is the 
model introduced by Peyrard and Bishop in 1989~\cite{PB89} and
later improved by Dauxois, Peyrard, and Bishop (DPB) to add a non-linear 
coupling between adjacent base pairs~\cite{DPB93}. 
In its original formulation a DNA configuration is described by a
set of continuous variables $y_i \geq 0$ giving the distance between
two base pairs along the sequence of length $N$ ($1 \leq i \leq N$).
The interaction between opposite bases is given by a short range Morse
potential and there is usually also a stacking interaction between
adjacent bases along the same strand.
A further refined version of the DPB model with helical degrees of
freedom was also studied~\cite{Barbi99,Barbi03}.

Many studies on the DPB model, in which specific types of dynamics were
employed, focused so far on equilibrium quantities, as the fraction of
open base pairs at a given temperature, or on dynamical aspects not
directly related to denaturation times.  A very recent study of the
dynamics of the DPB model used the reactive flux method
\cite{vanErp12}, a very powerful technique to deal with systems with
very slow timescales. In particular, the denaturation rate above the
melting temperature was investigated as a function of the sequence
composition and length. The rate was found to behave non-monotonically
as function of the sequence length and to be strongly influenced by the
sequence composition. From the data available in Ref. \cite{vanErp12}
it is however not possible to extract a dynamical exponent, therefore
the issue of the value of $z$ and of the relevance of helical degrees of
freedom in the different versions of the DPB model is currently still open.
Since the model describes bases regularly stacked, however, it would
seem not correct to utilize it for describing the chaotic entanglement
of denatureted DNA strands.

\section{Conclusions}

We have listed a variety of results concerning the scaling of DNA
denaturation time with its chain length. Each of the results can tell us
something about a particular aspect of DNA denaturation. In general, the
inclusion of helical degrees of freedom and consequently of geometrical
entanglement slows down the dynamics, giving rise to a higher exponent $z$.

We believe that the models of three-dimensional polymers are the most
suitable for describing full denaturation of very long DNA immersed
in a solvent above the denaturation temperature.  In this sense
it is a study of homopolymer dynamics, from an initial double helix
representing the double stranded DNA to a final disentangled state of two
polymers separated from each other. It would be interesting to measure
experimentally such disentangling time. As an indicator of polymers
separation one would need to consider some geometrical detectable aspect
and not a measure of chemical bonds between strands, which can break in
a  timescale much faster than the disentangling one.

\paragraph{Acknowledgments}

We acknowledge useful discussions with G.~Barkema,  M.~Peyrard,
F.~Sattin, T.~van~Erp, J.-C.~Walter, and with the participants of the
workshop {\em Inhomogeneous Random Systems} (Paris, January 2012).

\end{document}